\documentclass[pra,showpacs,floatfix,nofootinbib]{revtex4}
\usepackage{amsmath}
\usepackage{amssymb}
\usepackage{amsthm}
\usepackage{amscd}
\usepackage{amsfonts}
\usepackage[dvips]{graphicx}
\usepackage{latexsym}
\usepackage{mathrsfs}
\usepackage{bm}
\usepackage{subfigure}
\usepackage{color}
\newcommand{\cP}{\ensuremath{\mathcal{P}}}
\newcommand{\cT}{\ensuremath{\mathcal{T}}}

\begin{document}

\title{Revisiting the $\cP\cT$-symmetric Trimer: Bifurcations, 
Ghost States and
Associated Dynamics}

\author{K.\ Li}
\affiliation{Department of of Mathematics and Statistics, University of
Massachusetts, Amherst, MA 01003-9305, USA}

\author{P.\ G.\ Kevrekidis}
\affiliation{Department of of Mathematics and Statistics, University of
Massachusetts, Amherst, MA 01003-9305, USA}

\author{D.\ J.\ Frantzeskakis}
\affiliation{Department of Physics, University of Athens, Panepistimiopolis,
Zografos, Athens 157 84, Greece}

\author{C.\ E.\ R\"uter}
\affiliation{Faculty of Electrical Engineering, Helmut Schmidt University, 22043 Hamburg, Germany}

\author{D.\ Kip}
\affiliation{Faculty of Electrical Engineering, Helmut Schmidt University, 22043 Hamburg, Germany}

\begin{abstract}
In this paper,
we revisit one of the prototypical $\cP\cT$-symmetric
oligomers, namely the trimer. We find all the relevant branches of
``regular'' solutions and analyze the bifurcations and instabilities thereof. 
Our work generalizes the formulation that was
proposed recently in the case of 
dimers for the so-called
``ghost states'' of 
trimers, which we also identify and connect
to symmetry-breaking bifurcations from the regular states. 
We also examine the dynamics of 
unstable trimers, as well as those of the 
ghost states in the parametric 
regime where the latter are found to exist.
Finally, we present the current state of the art for
optical experiments in $\cP\cT$-symmetric trimers, as well
as experimental results in a gain-loss-gain three channel 
waveguide structure.
\end{abstract}

\maketitle

\section{Introduction}

The study of 
$\cP\cT$-symmetry in both linear
and nonlinear systems has received a continuously growing amount of attention
over the past 15 years. This effort originally stemmed from the
realm of quantum mechanics~\cite{R1,R2}, 
where $\cP\cT$-symmetric Hamiltonians were proposed as presenting a 
viable alternative -- also due to the fundamental
nature of the corresponding parity ($\cP$) and time-reversal ($\cT$) symmetries -- 
to the postulate of Hermiticity. Nevertheless, and while experimental
realizations in the quantum mechanical framework are still less clear,
a critical observation that significantly advanced the field was
made in the context of optics both theoretically~\cite{Muga}
and experimentally~\cite{R18,R20}. 
Indeed, in this field, since losses are abundant and controllable
gain is possible, 
an experimental synthesis of $\cP\cT$-symmetric Hamiltonians was 
demonstrated. 
This opened a new chapter in the
relevant investigations by enabling the interplay of linear
such $\cP\cT$-symmetric Hamiltonians with the effects of nonlinearity.
The latter is a feature ubiquitously present in such optical systems and
a theme of particular interest and complexity in its own right.
This, in turn, spearheaded not only additional experimental investigations
in optics~\cite{miri} and in electrical circuit analogues of such
systems~\cite{R21}, but also paved the way for numerous 
significant theoretical contributions on the subject. 
As a small sample among the many relevant topics, we highlight
the facilitation of unidirectional dynamics~\cite{kot1},
the analysis of the universality of the dynamics~\cite{kot2},
the exploration of symmetry breaking effects~\cite{R20,pgk}, the study of
switching of beams~\cite{sukh1}, of solitons~\cite{R30add3}, the
formation of symmetric and asymmetric bright
solitary waves~\cite{R30add1,R30add2}, of breathers~\cite{baras} 
and their stability~\cite{R30add4}, of dark solitons~\cite{vassos,BKM},
of vortices \cite{vassos}, as well as the
emergence of ghost states~\cite{R46a,R46b,R46} and the generalization of
such ideas into vortex type configurations~\cite{leykam},
$\cP\cT$-symmetric plaquettes~\cite{kaili} and higher-dimensional
media~\cite{ricardo}.

One of the themes of particular interest within these studies concerns 
the so-called $\cP\cT$-symmetric ``oligomers''. While 
most of the relevant attention was focused on
dimers~\cite{kot1,sukh1,R46a,R46b,R46} (arguably
due to the corresponding experimental explorations of~\cite{R18,R20,R21}), 
configurations with more sites, such as trimers~\cite{pgk,rspa} and
quadrimers~\cite{pgk,kaili,rspa,konozezy}, have also attracted
recent interest. In the present work, our aim is to revisit one of
these configurations, namely the $\cP\cT$-symmetric trimer. Both
the optical~\cite{R20} and the electrical~\cite{R21} implementation
of the corresponding dimer strongly suggest that the experimental realization 
of such a trimer system may be feasible.  
Hence, it is particularly interesting and relevant 
to fully explore the $\cP\cT$-symmetric trimer, 
and provide 
analytical results complementing 
the earlier findings of Ref.~\cite{pgk}, as well as to report on the
current state-of-the-art regarding a possible experimental realization
thereof in optics.

Our presentation will be structured as follows. In section II, we will
present the model and theoretical setup of the trimer. We will devise
analytical algebraic conditions that are relevant towards identifying
the full set of standing
wave solutions for this configuration. Importantly, in addition to
the more standard stationary solutions, we will also identify
the so-called ``ghost states'' of the model~\cite{R46a,R46b,R46}.
These are states that, remarkably, albeit solutions of the steady
state equations, due to their complex propagation constant, are
{\it not} genuine solutions of the original dynamical equations.
Nevertheless, as has been argued in the case of the dimer~\cite{R46},
these are waveforms of potential relevance in 
understanding the system's dynamics. In section III, we will 
present the corresponding numerical
results. In particular, we will seek both regular standing wave
states and ghost states, and will build a full state diagram as a function
of the gain/loss parameter $\gamma$ of our $\cP\cT$-symmetric trimer.
In addition to the existence properties of the obtained
solutions, we will consider their
stability (and potential instabilities/bifurcations) and, finally,
we will examine the system's dynamics, how the instabilities are
manifested, both in the case of the ``regular'' standing wave
solutions and in that of the ghost states identified herein. In 
section IV we discuss possible realizations of $\cP\cT$ symmetric optical 
systems (with a particular view towards trimers) 
and describe actual experimental limitations that have to be overcome.
Finally, in section V, we will summarize our findings and present our
conclusions, as well as some directions for future study.

\section{Model and Theoretical Setup}

The prototypical dynamical equations for the $\cP\cT$-symmetric
trimer model read~\cite{pgk}:
\begin{eqnarray}
i\dot{u}_1&=&-ku_2-|u_1|^2u_1-i\gamma u_1
\nonumber
\\
i\dot{u}_2&=&-k(u_1+u_3)-|u_2|^2u_2
\nonumber
\\
i\dot{u}_3&=&-ku_2-|u_3|^2u_3+i\gamma u_3.
\label{trimer1}
\end{eqnarray}
Here, $u_j(t)$ ($j \in \{1,2,3\}$) are complex amplitudes, 
dots denote differentiation with respect to the variable $t$  
(which is the propagation distance in the context of optics),  
while $k$ and $\gamma$ represent, respectively, the inter-site coupling 
and the strength of the $\cP\cT$-symmetric gain/loss parameter.
In the above equations it is assumed that the first site sustains
a loss at rate $\gamma$, while the third site sustains an equal
gain. The middle site suffers neither gain, nor loss.
Following the spirit of Refs.~\cite{pgk,rspa}, 
we start our analysis by seeking stationary solutions of Eqs.~(\ref{trimer1}) 
in the form $u_1=a \exp(i E t) $,
$u_2= b \exp(i E t)$ and $u_3=c \exp(i E t)$, where
$E$ represents the nonlinear eigenvalue parameter. This way, 
we obtain from Eqs.~(\ref{trimer1}) the following 
algebraic equations:
\begin{eqnarray}
E a&=&k b+|a|^2 a+i\gamma a,
\nonumber
\\
E b&=&k (a+c)+|b|^2b,
\nonumber
\\
E c&=&k b+|c|^2c-i\gamma c.
\label{trimer2}
\end{eqnarray}
Let us now use a polar representation of the three ``sites'', namely,
$a=A\exp(i\phi_a)$, $b=B\exp(i\phi_b)$, and $c=C\exp(i\phi_c)$. 
Then, from Eqs.~(\ref{trimer2}), one can immediately infer that $A=C$, i.e., the amplitudes
of the two ``side-sites'' of the trimer are equal. In addition,
the amplitude of the central site is given, as a function of $A$, by:
\begin{eqnarray}
B^2=\frac{E \pm \sqrt{E^2-8 A^2 (E-A^2)}}{2}.
\label{extra1}
\end{eqnarray}
In turn, the algebraic polynomial
equation for the squared amplitude of $A^2 \equiv x$ is given by
\begin{eqnarray}
x [\gamma^2 + (E-x)^2]^2 -k^2 E [\gamma^2 + (E-x)^2]
-2 k^4 x +2k^4 E =0
\label{extra2}
\end{eqnarray}
Once $A$ is determined from Eq.~(\ref{extra2}) and subsequently
$B$ from Eq.~(\ref{extra1}), then the two relative phases between
the three sites of the trimer have to satisfy:
\begin{eqnarray}
\sin(\phi_b-\phi_a)=-\sin(\phi_b-\phi_c)=-\frac{\gamma A}{k B}
\label{trimer5}
\\
\cos(\phi_a-\phi_b)=\cos(\phi_b-\phi_c)=\frac{EA-A^3}{k B}
\label{trimer6}
\end{eqnarray}
The above formulation provides [via Eqs.~(\ref{extra1})-(\ref{extra2})
and (\ref{trimer5})-(\ref{trimer6})] the full set of stationary
solutions of the trimer system, for given values of the coupling
strength $k$, nonlinear eigenvalue parameter $E$, and gain/loss
strength $\gamma$. In what follows in our numerical section below,
we will fix two of these parameters ($E$ and $k$) and 
vary $\gamma$ to explore the deviations from the Hamiltonian limit
of $\gamma=0$. As an important aside, let us note here that the 
global freedom of selecting a phase (due to the U$(1)$ invariance
of the model) can be used to choose $\phi_b=0$. Then, it is evident
that $\phi_c=-\phi_a$, which combined with the amplitude condition
$A=C$ implies that $u_3 = \bar{u}_1$, where the overbar denotes
complex conjugation. Clearly, this condition is in line
with the demands of $\cP\cT$-symmetry for our system.

As 
mentioned above, in addition to the regular stationary
solutions for which $E$ is real, one can seek additional solutions
with $E$ being complex, i.e., $E= \hat{E} \exp(i\phi_e)$
The resulting waveforms
are quite special in that they are solutions of the stationary
equations of motion~(\ref{trimer2}), yet they are {\it not} solutions
of the original dynamical evolution equations~(\ref{trimer1}),
because of the imaginary part of $E$. Such ``ghost state'' solutions have
recently been identified in the case of the $\cP\cT$-symmetric
dimer~\cite{R46a,R46b,R46} and have even been argued to play
a significant role in its corresponding dynamics therein. In the present
case of the trimer, to the best of our knowledge, they have not
been previously explored. Such ghost trimer states
 will satisfy the following algebraic conditions:
\begin{eqnarray}
\sin\phi_a &=& \frac{A \left(B^2+2 C^2\right) \gamma }{B \left(A^2+B^2+C^2\right) k} \\
\cos\phi_a &=& \frac{A (B-C) (B+C) \left(-A^2+B^2+C^2\right)}{B \left(-A^2+B^2-C^2\right) k} \\
\sin\phi_c &=& -\frac{\left(2 A^2+B^2\right) C \gamma }{B \left(A^2+B^2+C^2\right) k} \\
\cos\phi_c &=& \frac{\left(-A^2+B^2\right) C \left(A^2+B^2-C^2\right)}{B \left(-A^2+B^2-C^2\right) k} \\
\sin\phi_e &=& \frac{(A-C) (A+C) \gamma }{\left(A^2+B^2+C^2\right) \hat E} \\
\cos\phi_e &=& \frac{A^4-B^4+C^4}{\left(A^2-B^2+C^2\right) \hat E},
\end{eqnarray}
From these equations, the amplitudes
$A$, $B$, $C$ can be algebraically identified
by applying the identity $\sin^2\phi+\cos^2\phi=1$ for each of the
above angles. The relevant six algebraic
equations lead to the identification of the six unknowns, namely
the three amplitudes, as well as the phases $\phi_a$, $\phi_c$ and
$\phi_e$ (for simplicity we have set $\phi_b=0$ hereafter, 
without loss of generality). It should be noted here
that should such ghost state solutions be present with 
$\phi_e \neq 0$, these will spontaneously break the
$\cP\cT$ symmetry, given that they will have $A \neq C$.

Notice that for each branch of solutions that we identify
in what follows, we will also examine its linear stability. 
This will be done through a linearization ansatz of
the form $u_i = e^{i E t} [v_i + \epsilon 
(p_i e^{\lambda t} + \bar{q}_i e^{\bar{\lambda} t})]$. Here the $v_i$'s for
$i=1,2,3$ will denote the values of the field at the standing wave
equilibria,
while $\lambda$ are the corresponding eigenvalues and $(p_i,q_i)$
for $i=1,2,3$ denote the elements of the corresponding eigenvector
which satisfies the linearization problem at O$(\epsilon)$; the overbar
will be used to denote complex conjugation. When the eigenvalues
$\lambda$ of the resulting $6 \times 6$ linearized equations have
a positive real part, the solutions will be designated as 
unstable (whereas otherwise they will be expected to be dynamically stable).

We now turn to the detailed numerical analysis of the corresponding
stationary, as well as ghost branches of solutions.

\section{Numerical Results}

\begin{figure}[b]
\scalebox{0.5}{\includegraphics{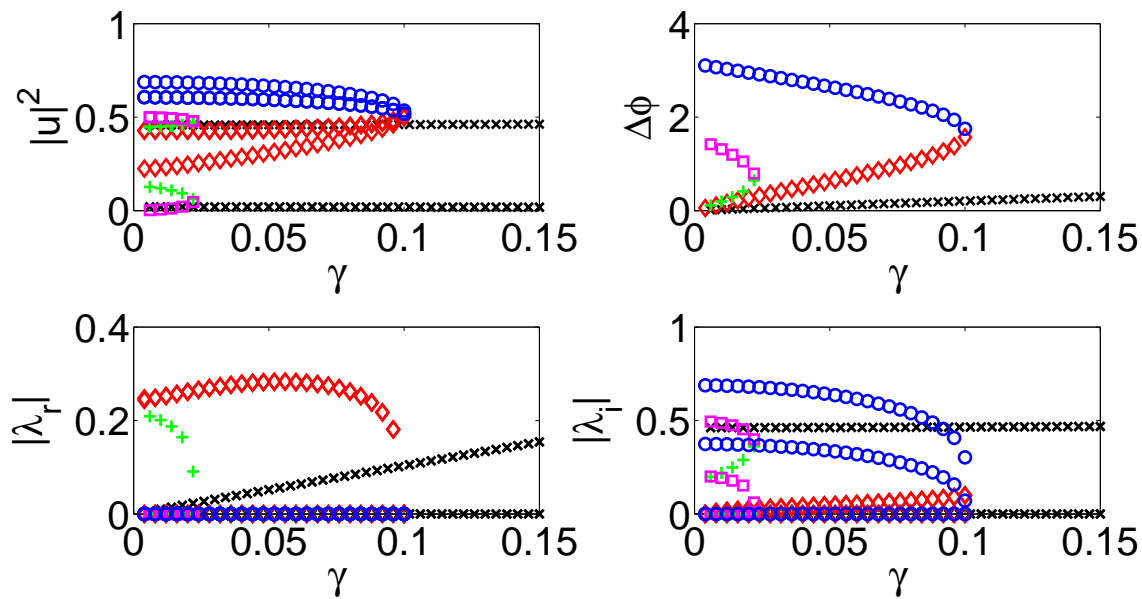}} %
\caption{(Color online) The solution profile of Eq.~(\ref{trimer1})
with $E=0.5$, 
$k=0.1$ and $\phi_b=0$. 
The four panels denote the solution amplitude
(top left), phase differences between adjacent nodes
(top right), real and imaginary
parts (second row) of eigenvalues.}
\label{trimer_e05k01}
\end{figure}

Since Eq.~(\ref{extra2}) is a polynomial of degree 5, we expect at 
most 5 distinct real roots (and at least 1 such). Indeed
for suitable choices of the free parameters $(E,k)$, 
we identify five branches of stationary
solutions. Figure~\ref{trimer_e05k01} illustrates a situation with
five branches under $E=0.5$ and $k=0.1$. Two of them, denoted by
blue circles and red diamonds, collide and terminate at $\gamma=0.1$.
The blue circles are essentially stable while the red diamonds are
unstable. Another pair of branches, namely the magenta squares
and green pluses collide and terminate at $\gamma=0.02$, with
the magenta squares being stable and the green pluses being unstable
(i.e., both of the above collisions are examples of 
saddle-center bifurcations).
The black crosses branch, which is essentially unstable,
persists beyond $\gamma=0.1$. 
Notice that the amplitudes of the different nodes for
this branch shown in the top left
panel of the figure are not constant: the upper line (standing for $B$) is slightly
increasing and the lower line (standing for $A=C$) is slightly decreasing.

\begin{figure}[tph]
\scalebox{0.5}{\includegraphics{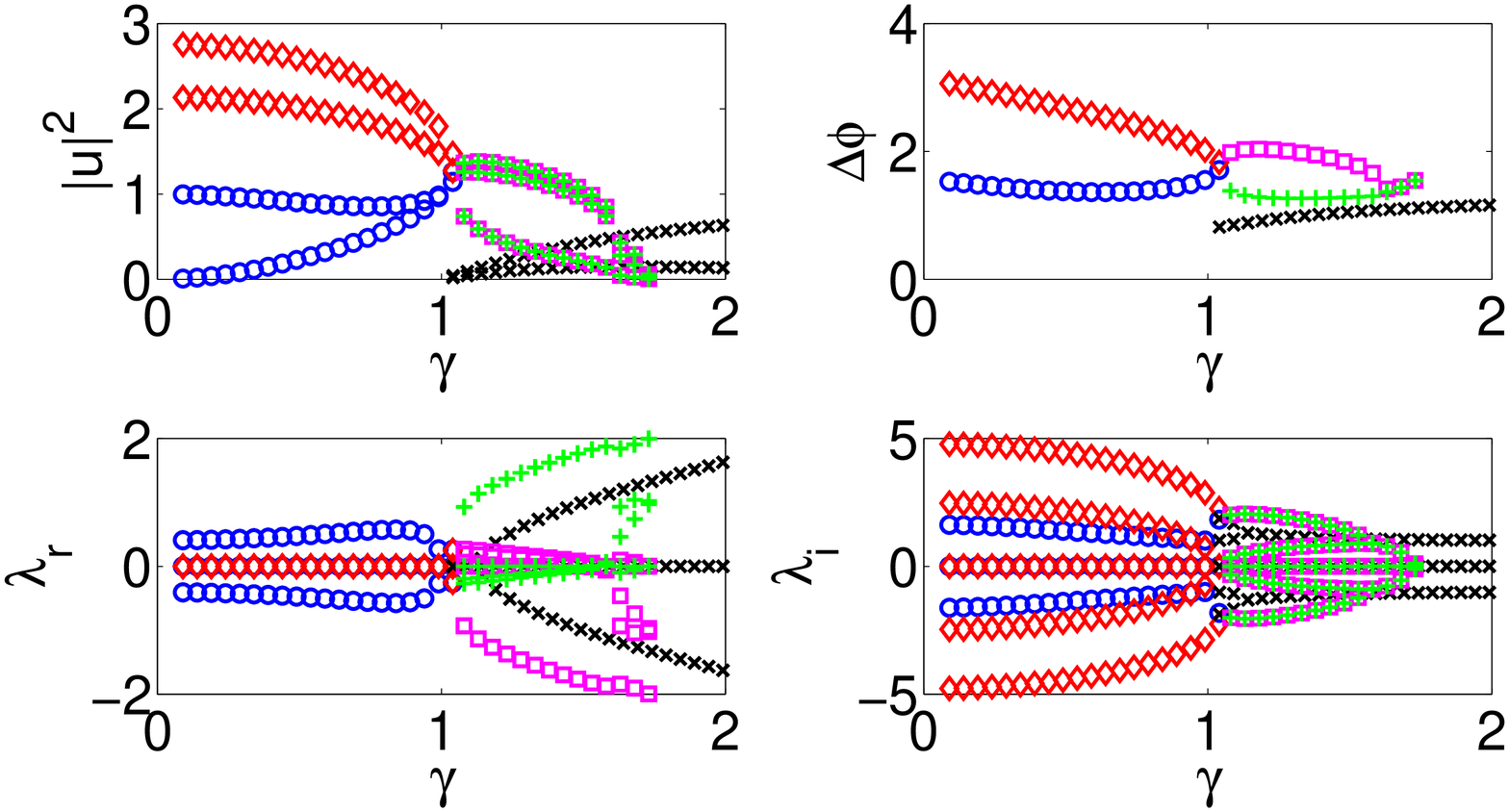}} %
\caption{(Color online) In a way similar to that of the previous
figure (i.e., with top left denoting amplitudes, top right relative phases,
bottom left real and bottom right imaginary part of the linearization
eigenvalues), the 4 panels show the existence and stability of solutions for 
a trimer with parameters $E=k=1$. There are three
regular standing wave branches:
the blue, the red and the black; the blue and red are the ones disappearing
hand-in-hand at $\gamma=1.043$.
Two ghost solutions are colored in magenta and green and bifurcate at the
destabilization of the blue branch for $\gamma=1.035$, while they
terminate for $\gamma=1.732$.}
\label{trimer}
\end{figure}

\begin{figure}[tph]
\subfigure[]{\scalebox{0.38}{\includegraphics{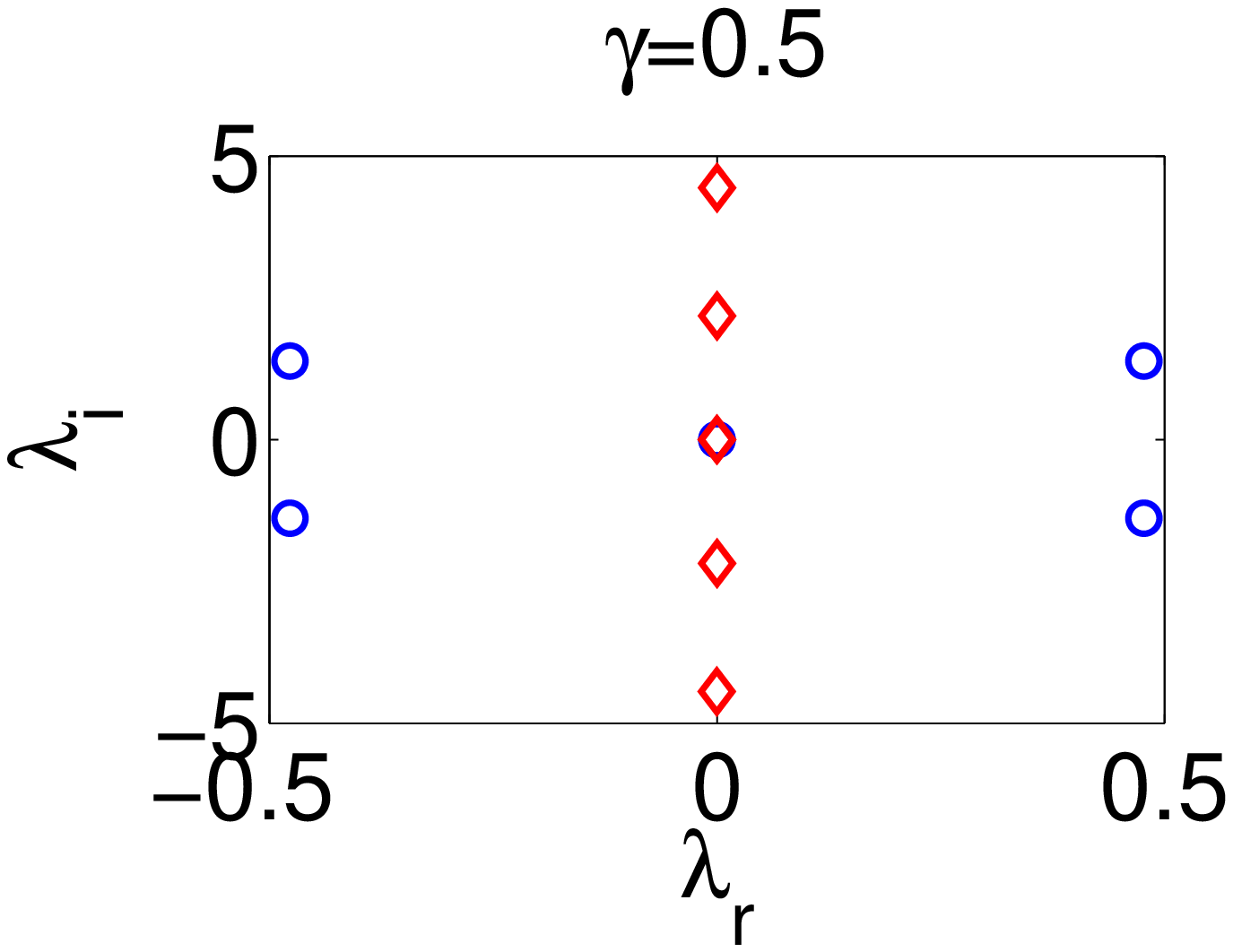}}}
\subfigure[]{\scalebox{0.38}{\includegraphics{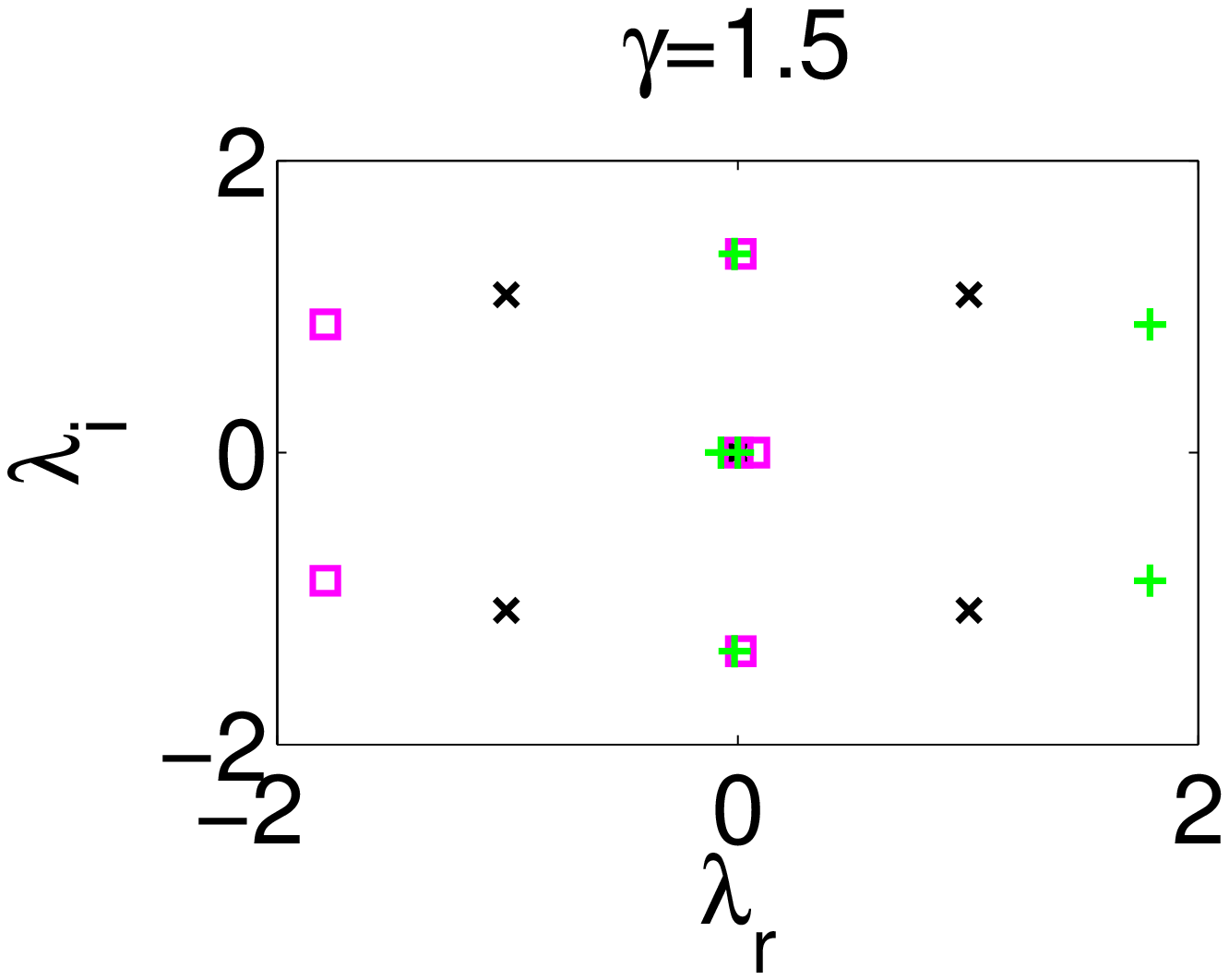}}}
\subfigure[]{\scalebox{0.38}{\includegraphics{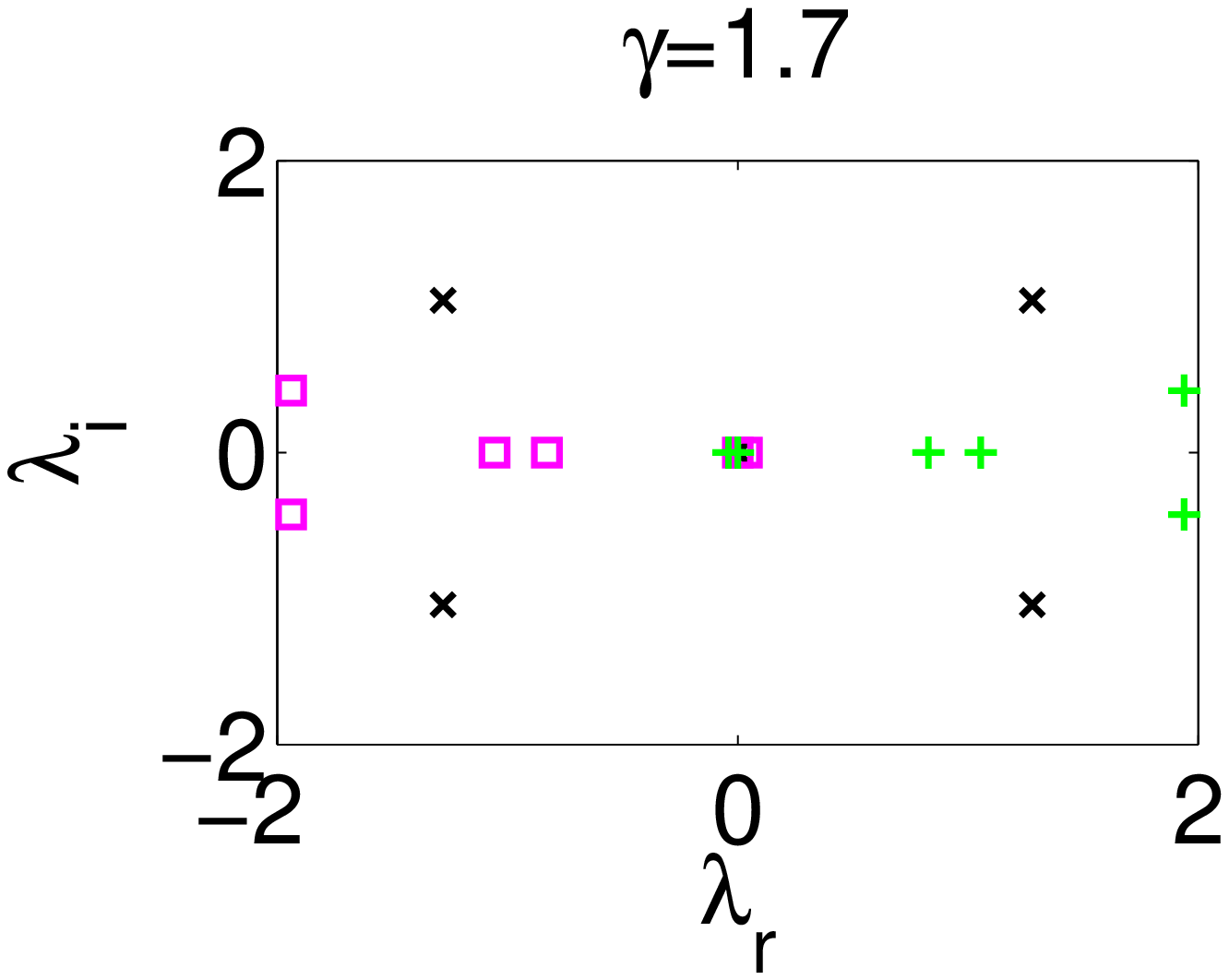}}}
\caption{(Color online) The spectral planes $(\lambda_r,\lambda_i)$
of the eigenvalues $\lambda=\lambda_r + i \lambda_i$ of the solutions shown in
Fig.~\ref{trimer}. The first panel shows the case of $\gamma=0.5$
where only the standing wave branches exist (blue circles -- unstable
and red diamonds -- stable). The second panel for
$\gamma=1.5$ has only one standing wave (black crosses -- unstable), and
two asymmetric ghost states which are mirror images of each other
(and so are their spectra), namely magenta squares and green pluses.
The third panel shows the same branches as in top right but for
$\gamma=1.7$ close to the termination of the ghost state branches.}
\label{trimer_stab}
\end{figure}

\begin{figure}[tph]
\subfigure[\ blue]{\scalebox{0.295}{\includegraphics{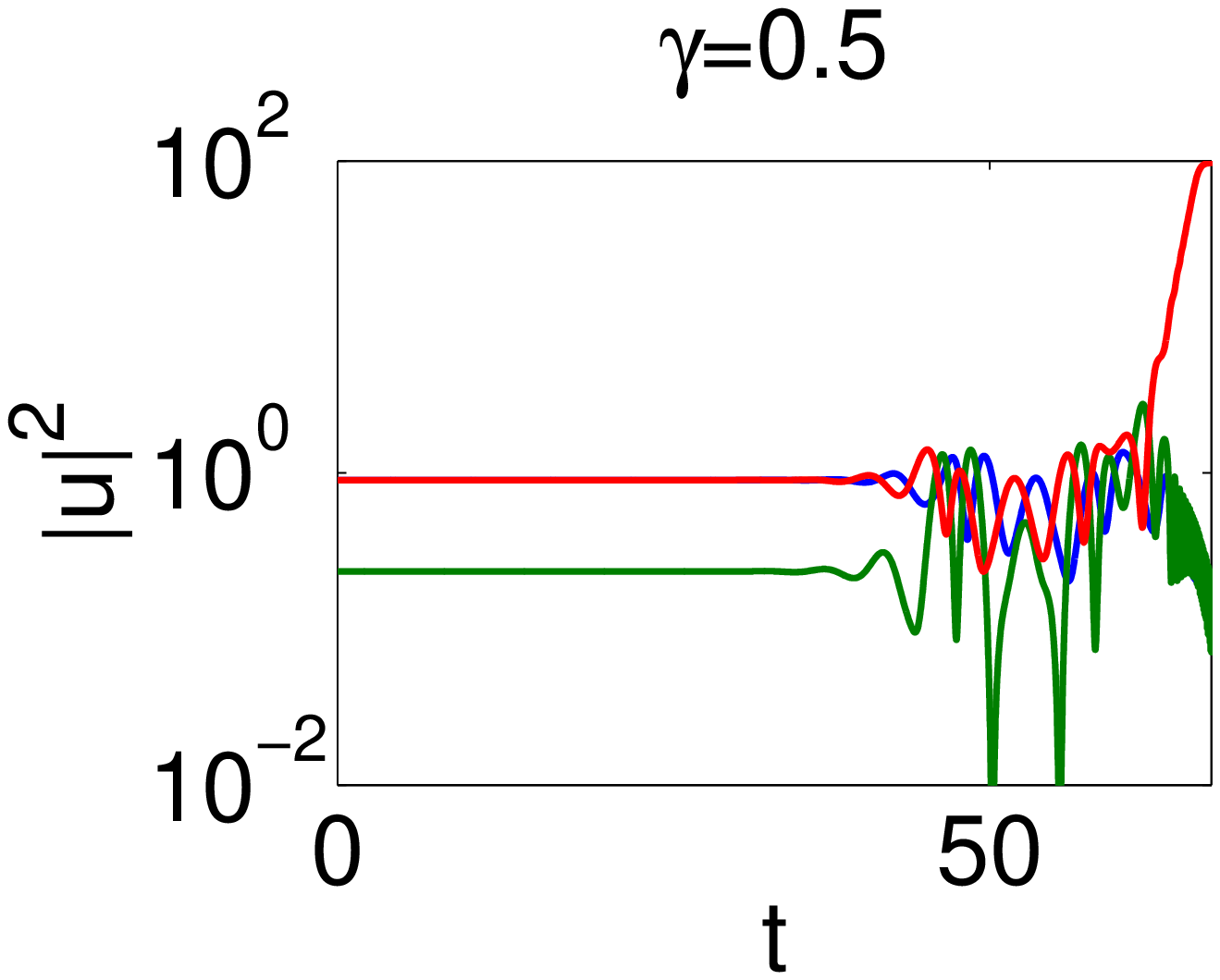}}}
\subfigure[\ blue circles]{\scalebox{0.295}{\includegraphics{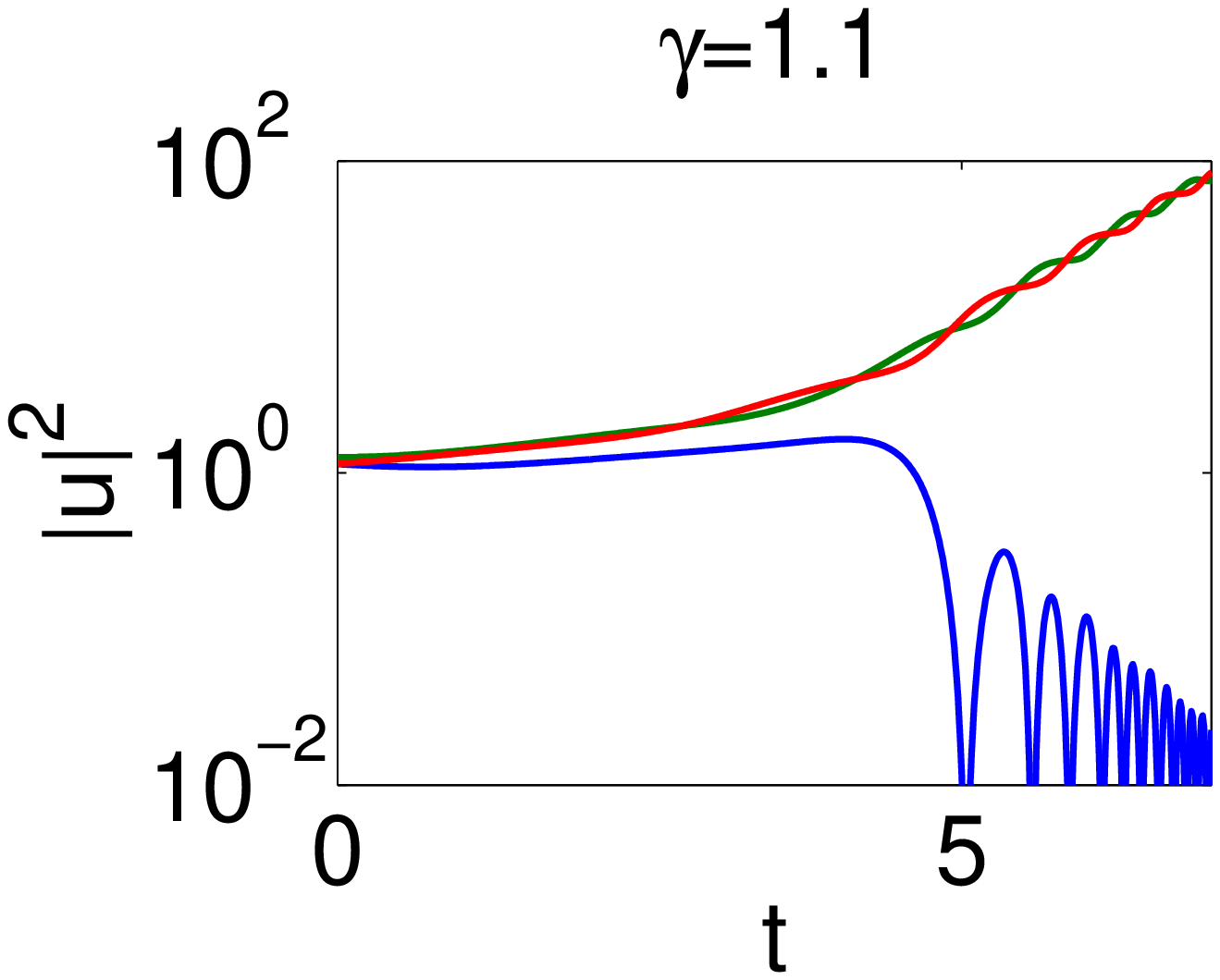}}}
\subfigure[\ red diamonds]{\scalebox{0.295}{\includegraphics{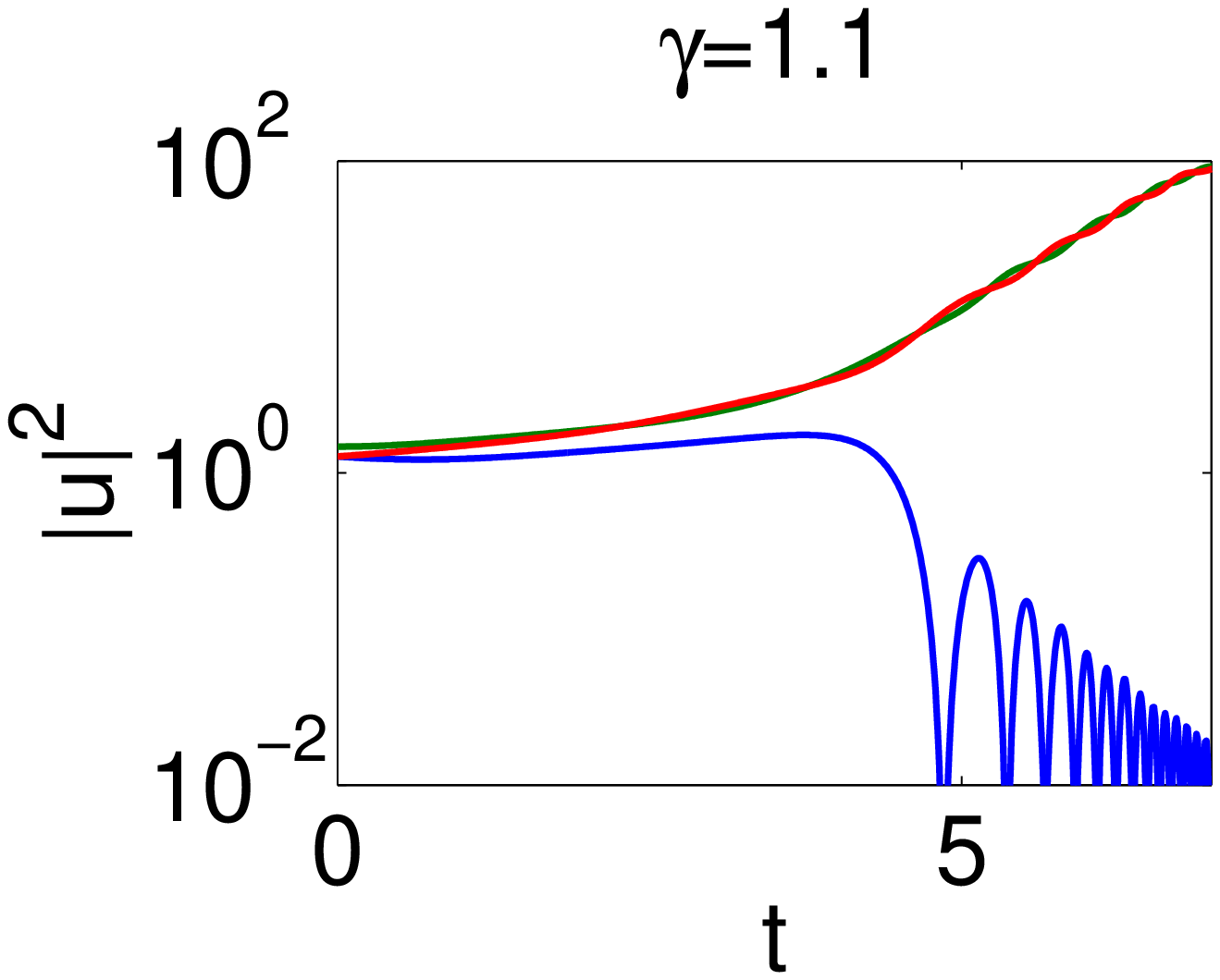}}}
\subfigure[\ black crosses]{\scalebox{0.295}{\includegraphics{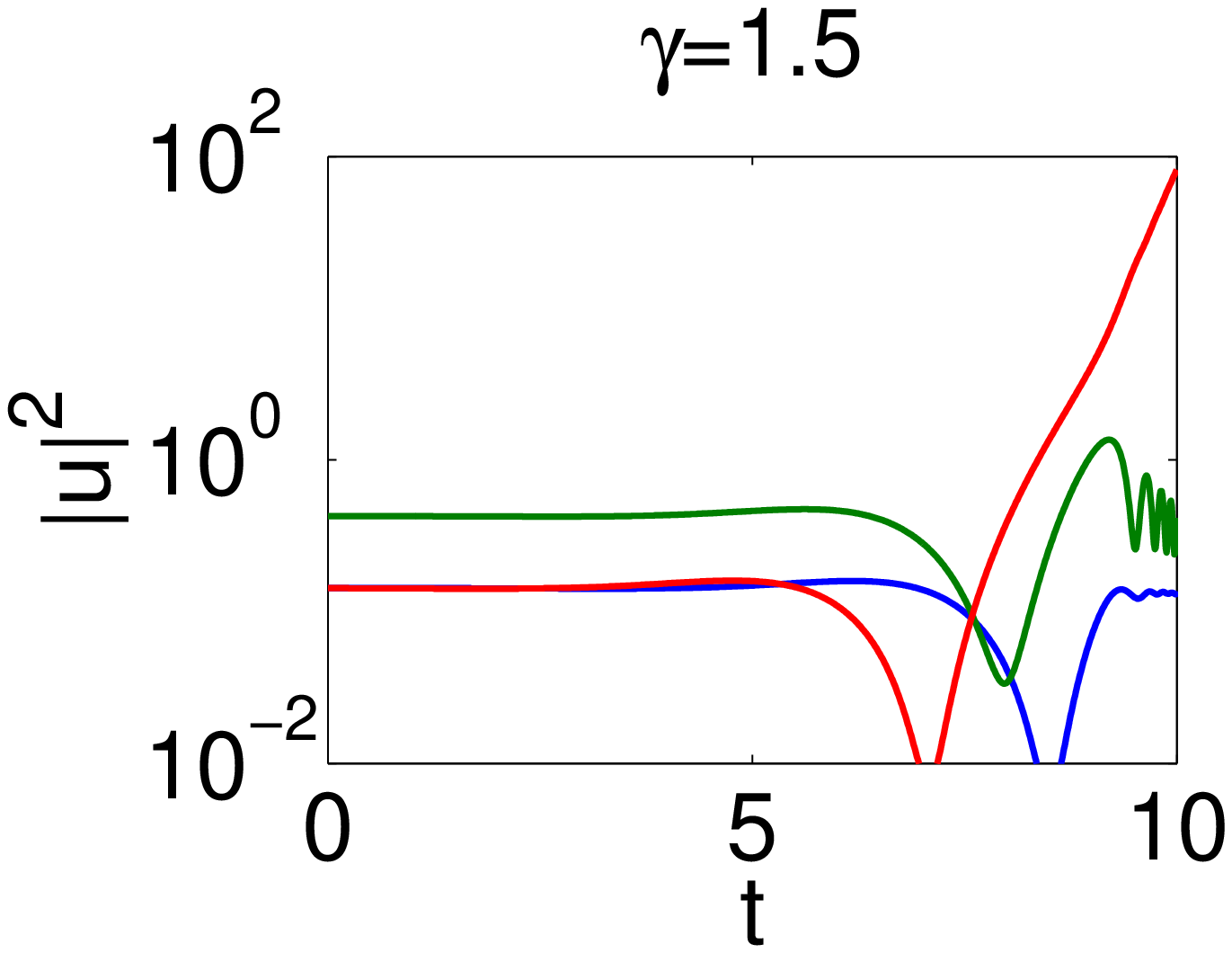}}} %
\subfigure[\ magenta squares]{\scalebox{0.295}{\includegraphics{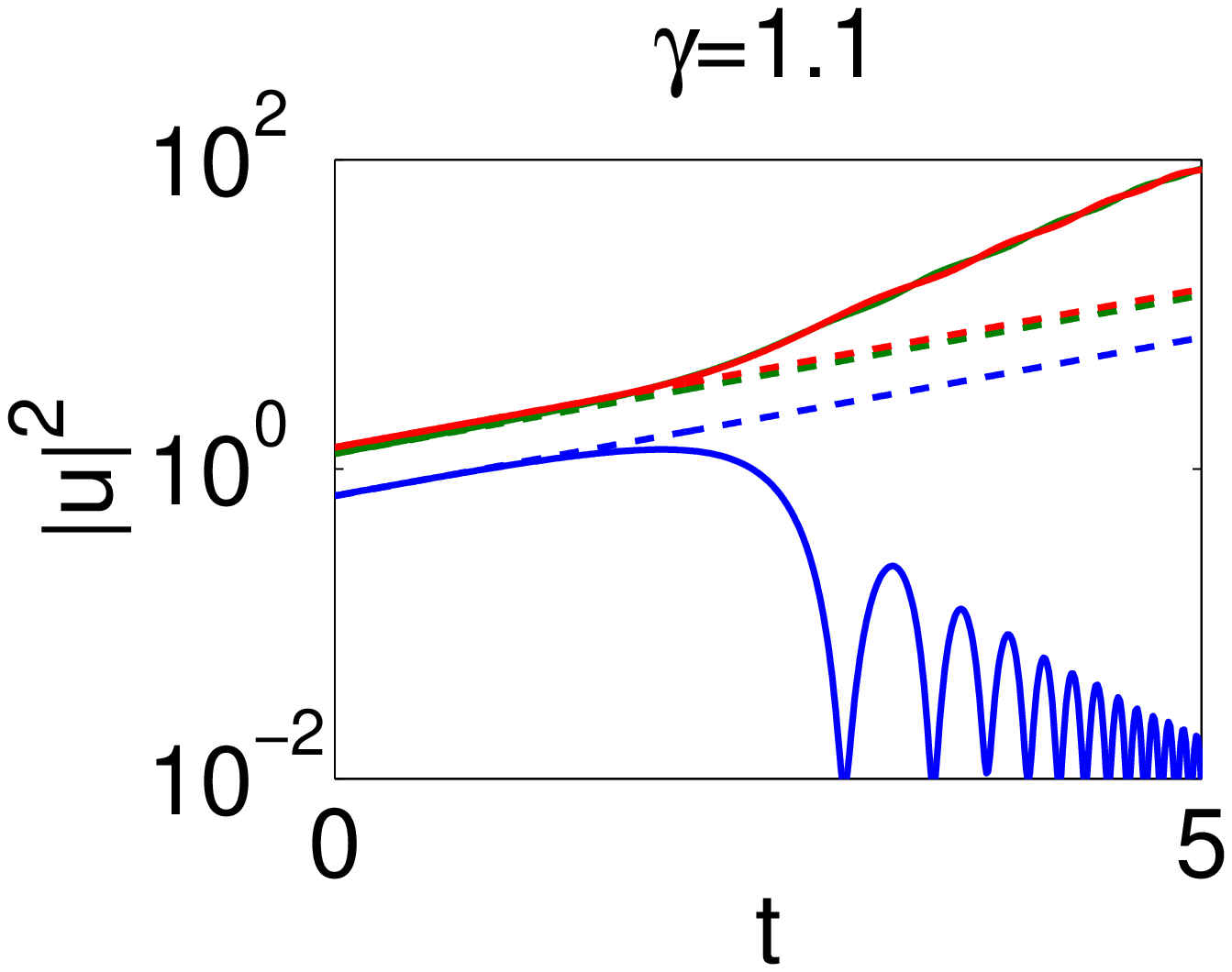}}}
\subfigure[\ magenta squares]{\scalebox{0.295}{\includegraphics{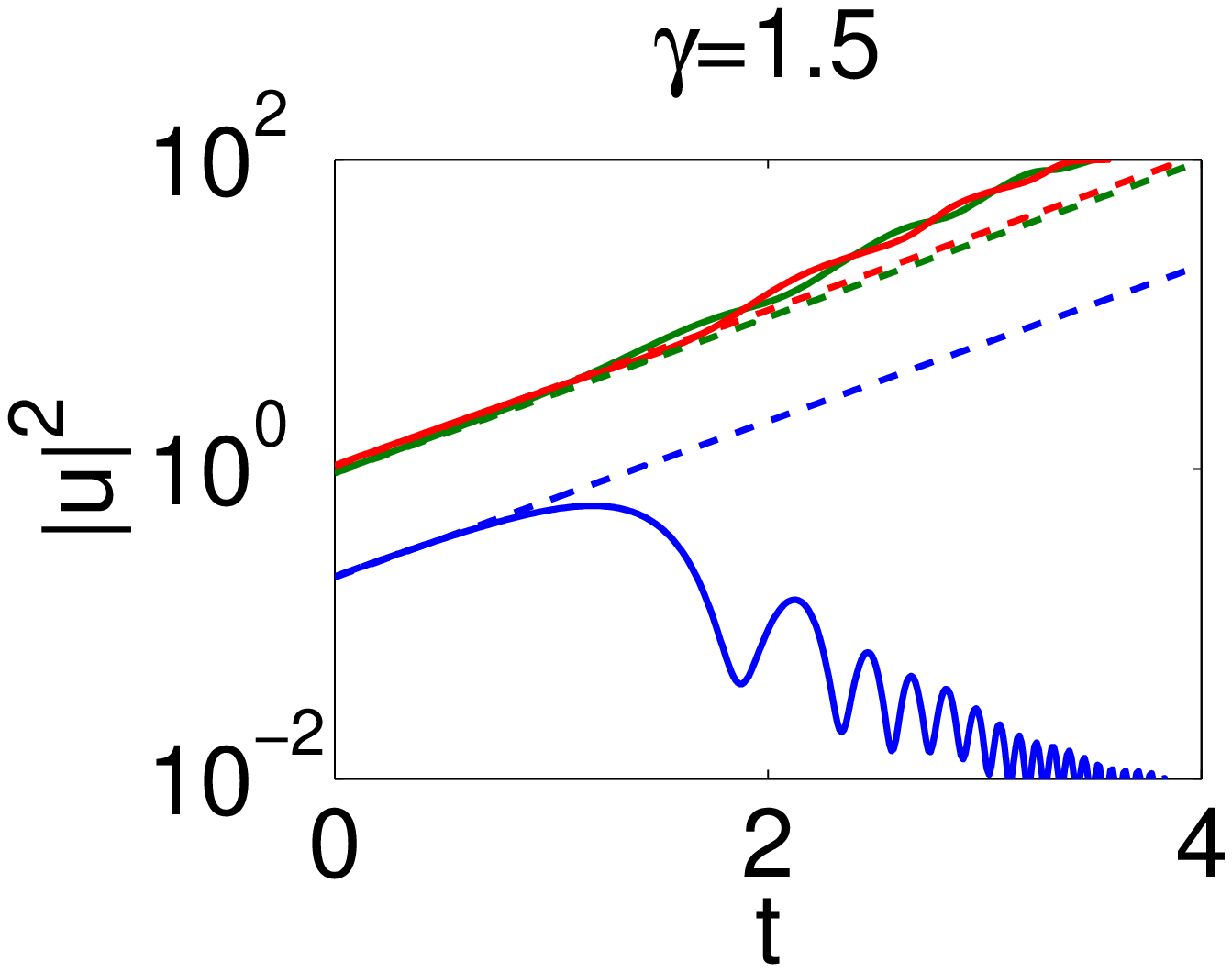}}} %
\subfigure[\ green pluses]{\scalebox{0.295}{\includegraphics{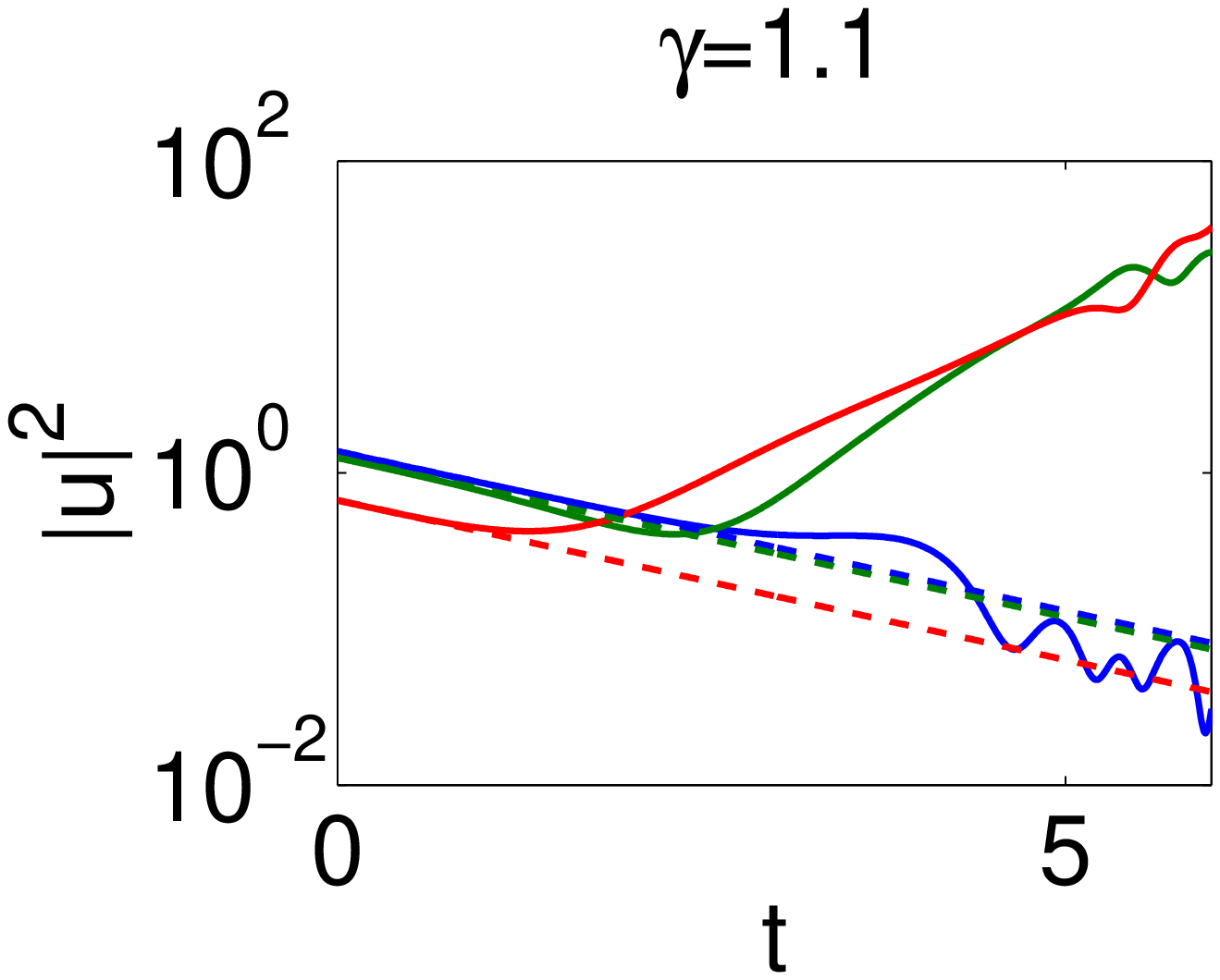}}}
\subfigure[\ green pluses]{\scalebox{0.295}{\includegraphics{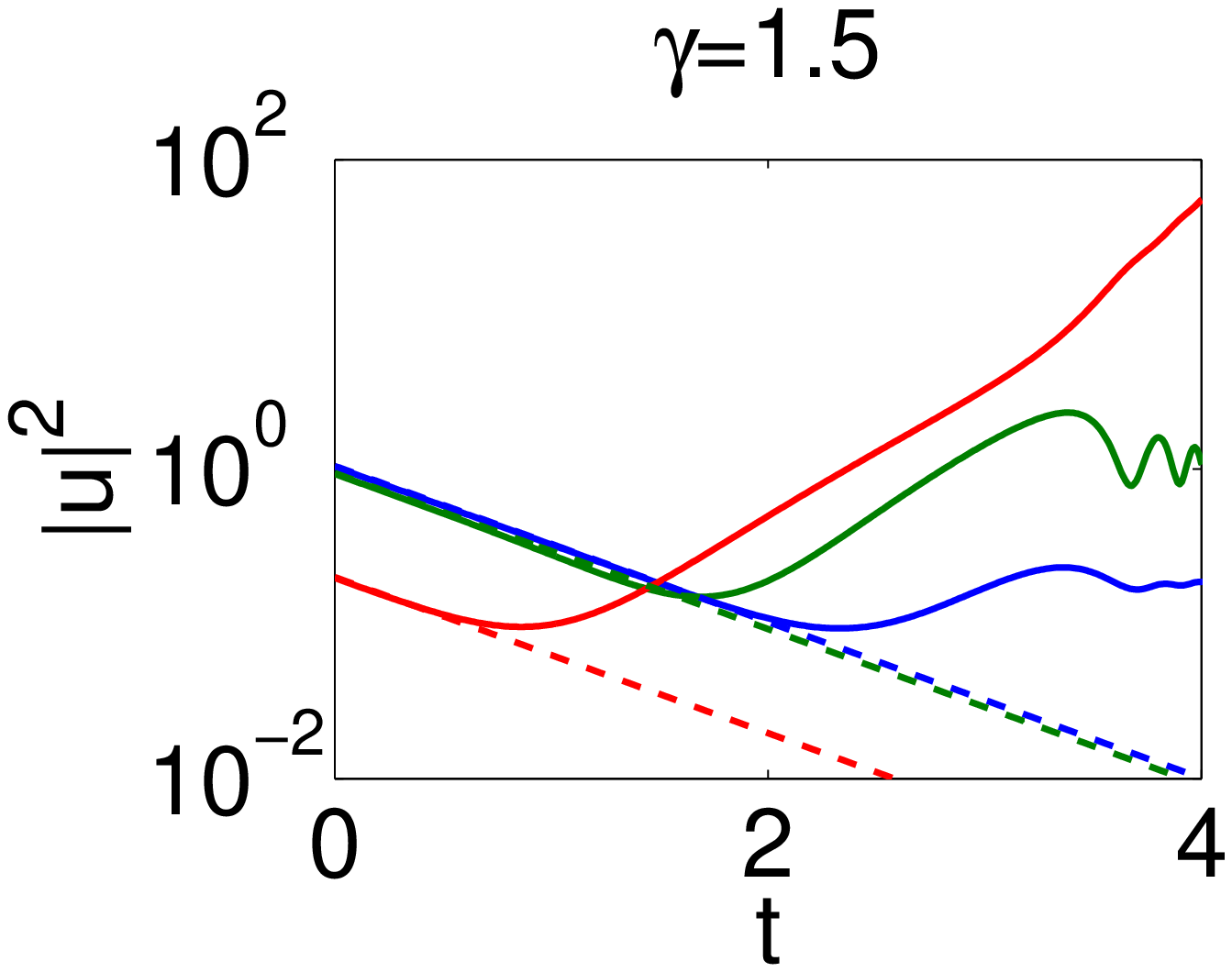}}}
\caption{(Color online) The dynamical evolution of the amplitudes of
the three sites for the solutions shown in Fig.~\ref{trimer}. Notice that
all solutions are plotted in semilog. The first row shows the evolution
of the three stationary branches. In (b) and (c), since these
branches are absent for $\gamma=1.1$, their profile for $\gamma=1.043$
is initialized. The second row shows dynamics of the two ghost 
state solution branches.
The dashed lines are the predicted dynamics of the ghost states
on the basis of their growth (for magenta squares) or decay (for
green pluses) rates.}
\label{trimer_dyn}
\end{figure}

In the following, 
we focus on a typical example of the branches
(both stationary and ghost ones) for a selection of the free
parameters of order unity, 
more specifically for
$E=k=1$; cf. Fig.~\ref{trimer}. We identify
three distinct examples of stationary states denoted by the blue circle,
red diamond and black cross branches. The blue circle and
red diamond branches stem from the corresponding ``+0--'' and ``--+--''
branches, respectively, namely the second and third excited
state of the Hamiltonian trimer problem of $\gamma=0$;
cf. with Ref.~\cite{todd}.
The blue circles branch is mostly unstable, except for
a small interval of $\gamma \in [1,1.035]$, while
the red diamonds branch
is chiefly stable, except for the narrow interval of values of
$\gamma \in [1.035,1.043]$.
In this narrow interval, the eigenvalues of both of these
branches are very close to each other.
For the blue circles branch, we also note that two eigenvalue pairs
stemming from a complex quartet collide on the imaginary axis
at $\gamma=1$ and split as imaginary thereafter. One of these
pairs exits as real for $\gamma>1.035$, and the two branches
(blue circles and red diamonds) collide shortly thereafter, i.e.,
at $\gamma=1.043$.


On the other hand, Fig.~\ref{trimer} also reveals an additional
branch, denoted by black crosses, that bifurcates from zero
at $\gamma=1$ and persists for all values of $\gamma$ thereafter.
This is quite interesting in its own right
as an observation since, as highlighted in Ref.~\cite{pgk}, the linear
critical point for the $\cP\cT$ phase transition is $\gamma=\sqrt{2} k$.
Thus, this branch presents the simplest oligomer example (ones such
are {\it absent} in the case of the dimer) whereby nonlinearity
enables a solution family to persist past the point of the linear limit
$\cP\cT$ phase transition. Additionally, it should be noted
that the branch is stable for all values of $\gamma < 1.13$,
but destablizes for all larger values of $\gamma$.

In Fig.~\ref{trimer}, however, in addition to the standard
stationary solutions, the ghost state solutions are also shown.
These are designated by the magenta squares and green pluses in the figure.
These ghost solutions are also
obtained for $\hat E=k=1$, and importantly (and contrary to what
is the case for the stationary states), they bear distinct amplitudes
in all three sites.
The two (magenta and the green) branches shown in the figure are mirror images
of each other, i.e., $A,B,C$ in the magenta branch are the
same as $C,B,A$ in the green branch, respectively,
and their phase difference and eigenvalues are opposite to each other.
Notice that as indicated above the difference in the magnitudes of
$A$ and $C$ supports the fact that these branches defy
the expectations of the  $\cP\cT$ symmetry. Indeed,
both of the branches arise through a symmetry-breaking bifurcation
 from the blue branch when it becomes
unstable at $\gamma=1.035$. Furthermore, it should be noted that the
branches terminate at vanishing amplitude
for $\gamma=1.732$. It is interesting to point out 
that when performing linear stability analysis of these states, we find
{\it both} of them to be unstable. Case examples of the linearization results
for both the regular states and the ghost ones are shown for three different
values of $\gamma$ in Fig.~\ref{trimer_stab}. For $\gamma=0.5$,
the red diamond branch is (marginally) stable, while
the blue circle branch bears the instability that we discussed
above for $\gamma < 1$. For $\gamma=1.5$, only the black branch
is present among the stationary ones and the magenta and green ghost
state branches manifest their respective asymmetries with spectra
that are {\it asymmetric} around the imaginary axis. This is a
characteristic feature of the ghost states; see also~\cite{R46,ricardo}.
Although among the two branches, the magenta is more stable 
and the green
highly unstable, even the magenta branch is predicted to be weakly
unstable with a small real positive eigenvalue. We will examine the
dynamical implications of these instabilities in what follows.
The last panel
similarly shows the case of $\gamma=1.7$ shortly before the disappearance
of the ghost state branches.

Finally, we 
examine the dynamics of the different
branches in Fig.~\ref{trimer_dyn}. The top row panels of the
figure show the evolution of the three stationary branches.  
Panels (a) and (b) show the blue circle branch for
$\gamma=0.5$ and $\gamma=1.1$, while panel (c) depicts the red diamond
standing wave branch for $\gamma=1.1$. Notice that the cases of
(b) and (c), the corresponding
branches cease to exist at $\gamma=1.043$. Thus in these
runs, we have used the terminal point profile of
the branches (at $\gamma=1.043$) 
as initial data for the evolution with $\gamma=1.1$.
Importantly, we note that in the unstable evolution of cases (b) and (c),
two of the sites end up growing indefinitely while the lossy
site ends up decaying. 
On the contrary, in the case (a), only the site with gain is led to
growth, while the other two are led to eventual decay.
In panel (d), we show the black crosses
branch, the third among the standing wave solutions
identified herein for $\gamma=1.5$.
Notice that panel (d) shows a different dynamical evolution from 
panels (b) and (c) and more in line with panel (a), 
showcasing that there are indeed two general growth scenaria: one
in which the gain site ``grabs'' along the neutral central
site and leads it to indefinite growth and one in which the
central site is ultimately led to decay together with the lossy
site. 

The four panels in the lower row show the dynamical evolution of the
two ghost states (green pluses and magenta squares) for the cases of 
$\gamma=1.1$ and $\gamma=1.5$. 
With $E=E_r + i E_i$ being complex, the ghost state solutions
under the form $u_1=\exp(i E t) a$,
$u_2=\exp(i E t) b$ and $u_3=\exp(i E t) c$ should evolve exponentially,
as indicated by dashed lines in panels (e)-(h). 
In particular, the magenta squares branch with negative $E_i$ is 
expected to lead to growth (for all three nodes of the trimer), 
while the green plus branch with positive
$E_i$ is anticipated to decay (again for all nodes). 
The slopes of these
growth/decay features 
are given by $-2E_i=-2\hat E\sin\phi_e$. 
However, in line with their anticipated linear ``instability'',
neither of these follows exactly the
dynamics anticipated above. Both of them evolve for a short period
according to the expected growth or decay, and
then the gain sites start to grow and the loss sites start to decay,
regardless of the trend predicted by the form of the ghost state
(discussed above).
Moreover, it is relevant to note as regards the
corresponding dynamics that the cases of the blue circle and red
diamond branches of $\gamma=1.1$ exhibit similar (asymptotic) dynamics to those
of the magenta squares and of the green pluses for the same
parameter value; i.e., the central site is also led to growth along
with the gain one. On the other hand, it is also evident that the black crosses branch for
$\gamma=1.5$ instead follow an evolution resembling to the 
asymptotic evolution of the green pluses branch (which is different
from that of the magenta squares for the latter value). I.e., here
only the gain site is ultimately led to growth.

Having unveiled the existence and stability, as well as nonlinear
dynamical characteristics
of the different solutions, we now turn to a discussion of the potential for
experimental realization of $\cP\cT$-symmetry in optical systems
especially as regards trimers, but also more generally.

\section{Experimental realization of $\cP\cT$ optical systems}

General requirements for realization of $\cP\cT$ optical systems are the availability of adequate methods for formation of coupled waveguide systems or arrays with the additional opportunity to spatially tailor loss and gain in the substrate. In other words, suitable fabrication conditions should allow for spatial manipulation of both real and imaginary part of the dielectric constant. 

Laser crystals and glasses are amplifying media that may provide the necessary optical gain by using different physical mechanisms. Examples are doped bulk crystals and fibers that make use of stimulated emission to amplify a weak signal, electron-hole recombination in semiconductor optical amplifiers (SOAs), parametric amplification in nonlinear crystals, or stimulated Raman scattering (SRS). Furthermore, many laser gain materials allow for the formation of guiding index structures. However, although many amplifying materials exist, the most challenging aspect is the necessity to achieve optical gain while at the same time the real part of the refractive index should remain constant (or change only in a negligible way in order to maintain $\cP\cT$ symmetry), which is difficult to achieve because of the Kramers-Kronig relation. Besides thermo-optic effects in case of high optical powers, it turns out that other mechanisms like self-and cross-phase modulation are limiting the suitability of most laser gain media for application in $\cP\cT$ optics.

Another well-known amplification mechanism is optical beam coupling in photorefractive media like photovoltaic lithium niobate (LiNbO$_{3}$) crystals, which exists already at quite low optical light powers. Due to advanced waveguide formation techniques, LiNbO$_{3}$ is a favorite material for use in integrated optics \cite{Kip_APB1998}. Besides diffraction of weak signal beams on recorded index gratings which leads to gain, the point symmetry $3m$ of LiNbO$_{3}$ enables interaction of orthogonally polarized waves too \cite{Novikov1987}. A polarization grating recorded by a pump and signal beam having mutual orthogonal polarization allows for optical signal gain; the small-signal gain can reach several tens per cm for strong Fe doping \cite{Novikov1987}. At the same time, the spatially varying polarization pattern of pump and signal beam does not induce significant phase changes for the interacting beams. Using this mechanism, the first experimental demonstration of $\cP\cT$ symmetry in optics has been achieved in Fe:LiNbO$_{3}$ using Ti in-diffusion to form coupled waveguide structures \cite{R20}. However, there exist still some limitations that have to be overcome in order to realize more advanced $\cP\cT$ symmetric optical settings.     

While spatial tailoring of optical gain may be achieved by limiting an optical pump beam to certain waveguide channels, this can be hardly done for loss (a technologically quite challenging solution consists in the formation of metallic stripes of precisely defined width on certain channels, see \cite{R18}). Due to such difficulties, a more realistic experimental approach consists in allowing for a constant loss in all coupled channels, while this loss is overcompensated by adjustable gain in some selected channels only. 

\begin{figure}[tph]
\includegraphics{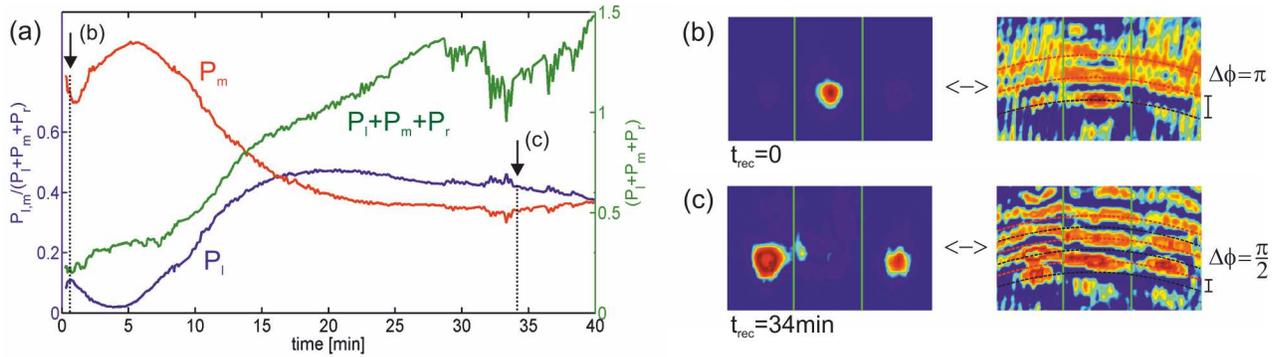}
\caption{(Color online) (a) Evolution of powers in a three-channel coupler. Left and right channels experience equal gain; the lossy central channel is excited from the input facet. Gain develops due to holographic recording of a polarization grating according to $\gamma(t_{rec})=\gamma_{0}(1-\exp(-t_ {rec}/\tau))$ with the photorefractive (Maxwell) time constant $\tau$. (b,c) Power distributions (left panel) on the end-facets for different recording times (b: $t_{rec}=0$; c: $t_{rec}=34$\,min) and corresponding interferograms (right panels) showing the phase relation of central and outer channels.}
\label{X1}
\end{figure}

\begin{figure}[tph]
\includegraphics{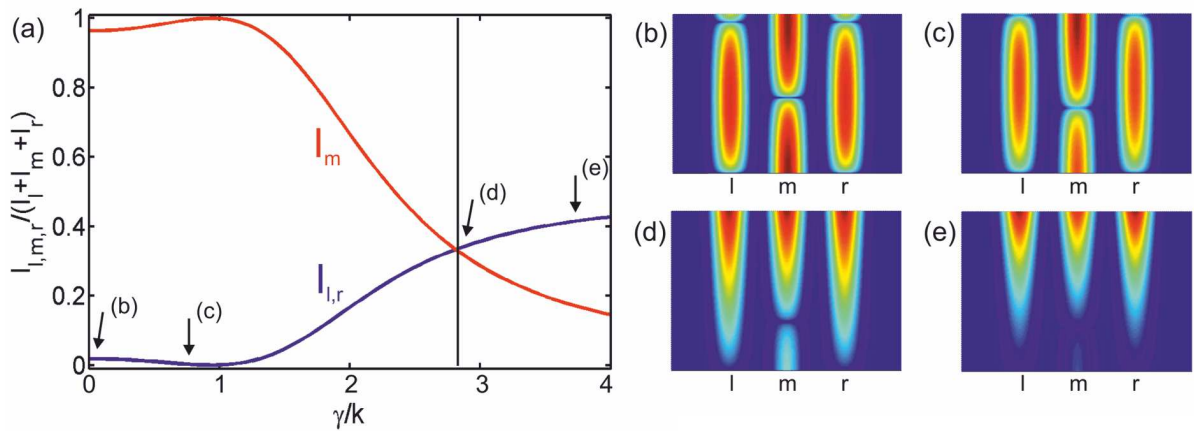}
\caption{(Color online) (a) Simulation of normalized intensity of the three-channel coupler as a function of gain (normalized to $k$) using integration of the coupled-wave equations. The vertical line corresponds to  $\gamma/k=2\sqrt{2}$, which is the 
``break even'' point of this gain-loss-gain system. The four panels (b-e) on the right show the light power evolution during propagation for $\gamma/k=0$, (b), $\gamma/k=0.7$ (c), $\gamma/k=2.8$ (d), and $\gamma/k=3.7$ (e). For the false-color scale we choose a proportionality to $\sqrt[4]{I}$ to improve visibility of the amplified signal light in the channels.}
\label{X2}
\end{figure}

An experimental example of a three-channel coupled waveguide structure having distributed gain and loss is shown in Fig.~\ref{X1}. 
While the precisely
 $\cP\cT$-symmetric pattern of loss-(neither gain nor loss)-gain is
not directly realizable in the above described experiments, here we
focus on a somewhat different configuration featuring an 
alternation of gain-loss-gain which is, arguably, the nearest
experimentally realizable optical variant. To achieve that,
three parallel single-mode waveguide channels for a wavelength of $\lambda=532\,$nm are formed on a Fe-doped x-cut LiNbO$_{3}$ substrate by in-diffusion of a stripe-like Ti film. The propagation length is 20\,mm and the linear coupling coefficient is $k\approx0.2\,$mm$^{-1}$. In the sample overall but constant loss is due to absorption of the used green light by in-diffused Fe ions. Similar to the arrangement in \cite{R20}, optical gain for the extraordinarily polarized signal is achieved by pumping the sample from the top using a plane wave of ordinary polarization. An amplitude mask on top of the substrate shields the central channel ($\sharp$2), thus (equal) gain is obtained for the left ($\sharp$1) and right ($\sharp$3) channels only. The signal light is coupled from the end-facet into the central channel. As can be seen, when the pump beam is switched on at time $t_{rec}=0$, power in the two outer channels start to increase. Simultaneously the total power (black symbols in Fig.~\ref{X1}) increases due to buildup of the polarization grating. Beside some asymmetries in the temporal evolution discussed below, for longer recording symmetry improves again, and the corresponding interferograms on the $rhs$ show the development of relative phase of central and outer channels, starting from the in-phase condition at $t_{rec}=0$ (Fig.~\ref{X1}b) towards a final phase difference of $\pm\pi/2$ at $t_{rec}\approx 34$\,min (Fig.~\ref{X1}c). This relative phase development is 
in line with the earlier theoretical expectations on the basis of
Eqns. such as~(\ref{trimer5})-(\ref{trimer6}).
This behavior is also in good agreement with simulations of this 
gain-loss-gain system based on Runge-Kutta integration of the corresponding coupled-wave equations in Fig.~\ref{X2}. Of course, these runs 
also manifest the partial
differences of this experimental realization from the genuinely
 $\cP\cT$-symmetric case in that ultimately all three waveguides feature
growing optical power in the simulations of Fig.~\ref{X2}, a trait
which is absent e.g. in Fig.~\ref{trimer_dyn} (where at least one
waveguide is not growing indefinitely in power).

Obviously, some experimental problems still exist. The temporal evolution in the left and right channels is far from being perfectly symmetric, especially for intermediate recording times. In most experiments, for long recording times (i.e.\ high gain) output powers of the three channels start to fluctuate slightly. Possible explanations for this behavior are large induced space-charge fields that may lead to spark plugs across the sample surface, or weak phase instabilities of the setup. A general problem is the small achievable gain, which is rarely sufficient to reach typical $\cP\cT$ symmetry-breaking thresholds in most of the fabricated samples. Gain is also limited because of low powers of signal and pump light: Higher signal power would lead to decoupling of the excited channel due to nonlinear index changes, while higher pump power would record a distorting phase gradient at the boundaries of the used amplitude mask (i.e. between illuminated/non-illuminated channels). 

For future experiments using Fe:LiNbO$_{3}$ waveguide samples, a main objective will thus be to increase optical gain by optimizing material properties. However, when doping LiNbO$_{3}$ substrates with Fe using in-diffusion, the physical mechanisms that may lead to high gain when coupling orthogonally polarized waves are not yet fully understood: The influence of certain diffusion atmospheres, interference from simultaneous Ti in-diffusing used for waveguide formation, or the effect of Li out-diffusion at high temperatures and consequences on possible lattice sites of in-diffused Fe ions have not been investigated in detail. In particular, the high gain found in some Fe-doped bulk LiNbO$_{3}$ crystals has not been observed in waveguide samples so far.

An alternative experimental $\cP\cT$-symmetric model system that also uses LiNbO$_{3}$ with its well-developed waveguide fabrication technology as a substrate is Er doping to achieve gain in the optical communication window at $1.5\,\mu$m. Although no detailed data on cross-phase modulation when pumped e.g.\ with 980\,nm wavelength is available yet, data from Er-doped fiber amplifiers (EDFAs) indicates that induced phase changes might be sufficiently small \cite{Jarabo1997}. Work on such systems is currently in progress, which may perhaps allow avoiding the described unwanted nonlinear effects that disturb the symmetry condition of the (real) refractive index in Fe-doped photorefractive LiNbO$_{3}$ for higher input powers.

\section{Conclusions \& Future Challenges}

In the present work, we have revisited the theme of one of the
prototypical $\cP\cT$-symmetric oligomers, namely the trimer.
We have illustrated the different number of branches 
(at least one and at most five) of standing wave
solutions that exist for
this system. We have thereafter focused on a case example of parameters
of order unity and have shown that two of these standing wave
branches terminate in a pairwise disappearance, while the third
one exists for values of the gain/loss parameter, in fact,
extending past the point of the $\cP\cT$-symmetry breaking
phase transition of the linear limit occuring at $\gamma=\sqrt{2} k$.

Additionally, we have also presented the formulation of
the so-called ghost states in this system and have explicitly
computed them, showing how they emerge through a symmetry breaking
bifurcation (asymmetrizing the amplitudes of the two side sites
$A$ and $C$) from one of the standing wave branches. As expected
on the basis of such an effective pitchfork bifurcation, the
two resulting ghost state branches are mirror-images of each
other (and so are their corresponding spectra). 
The dynamics of both the unstable stationary states and
those of the ghost states revealed two possible dynamical
scenaria. In one of these,
the ``neutral'' site (without gain or less) sided with the
gain one, while the other corresponded to the case where
it sided with the lossy site.

Additionally, we have explored the possibility of creating
$\cP\cT$ symmetric systems in nonlinear optics, revealing
that it is arguably simpler to create e.g. a gain-loss-gain
three-channel system, rather than the genuinely $\cP\cT$ symmetric situation
where a waveguide with gain and one with loss straddle a middle
one without either gain or loss. On the other hand, for this
gain-loss-gain setting, we presented both physical experiments and
corroborating numerical simulations revealing the partition
of the fraction of optical power (initially placed at the
central channel) and how it transfers more
to the outer gain channels as the gain is increased beyond $\gamma/k=1$.

There are 
many interesting questions that arise from the
present study and are worthy of further exploration.
It would be
interesting to generalize our considerations herein to the
case of quadrimers and to appreciate how the complexity
of the relevant configurations expands, especially since
in the latter case, there is generally the potential
of two gain/loss parameters~\cite{konozezy}; nevertheless
per the above discussion on experimental possibilities
in optics, the case of two waveguides with equal gain and two
with equal loss would appear as the most realistic one presently.
At the same time, further 
experimental implementations of oligomer systems, either
at the electrical circuit level, or at the optical waveguide level 
discussed in the last section would be particularly desirable
and highly interesting towards an increased understanding of
the systems' dynamics. Efforts in these directions are currently
in progress and will be reported in future publications.

\vspace{5mm}

\acknowledgments 
The work of P.G.K. is partially supported by the US
National Science Foundation under grants NSF-DMS-0806762, NSF-CMMI-1000337,
by the US AFOSR under grant FA9550-12-1-0332, as well as the
Binational Science Foundation under grant 2010239 and the
Alexander von Humboldt Foundation.
D.K. thanks the German Research Foundation (DFG, Grant
KI 482/14-1) for financial support of this research. D.J.F. is partially supported by the
Special Research Account of the University of Athens.

\end{document}